\documentstyle[12pt]{article}
\newlength{\dinwidth}
\newlength{\dinmargin}
\def\sel{\sigma_{el}^{VN}}
 \def\sin{\sigma_{in}^{VN}}
 \def\stot{\sigma_{tot}^{VN}}
 \def\inf{\int_{-\infty}^{\infty}}
 \newcommand\noi{\noindent}
 \newcommand\beq{\begin{equation}}
 \newcommand\eeq{\end{equation}}
 \newcommand\beqn{\begin{eqnarray}}
 \newcommand\eeqn{\end{eqnarray}}
 \newcommand{\doublespace} {
 \renewcommand{\baselinestretch} {1.6}
 \large\normalsize}

 \newcommand{\ra}{\rangle}
 \def\sel{\sigma_{el}^{VN}}
 \def\inf{\int_{-\infty}^{\infty}}

\begin{document}
\vspace*{1cm}
\hspace*{9cm}{\Large MPI H-V27-1996}
\vspace*{3cm}

\begin{center}
{
\renewcommand{\thefootnote}{\fnsymbol{footnote}}

{\Large {\bf Towards Study of Color Transparency\\
with Medium Energy Electron Beams}\footnote{
To appear in the Proceedings of the Workshop on Future Physics at HERA,\\
DESY, September 25, 1995 -- May 31, 1996}}\\
}
\vspace*{5mm}
{\large J\"org~H\"ufner$^a$ and Boris~Kopeliovich$^{bc}$}\\
\end{center}
$^a$ Inst. f\"ur Theor. Physik der Universit\"at ,
 Philosophenweg 19, 69120 Heidelberg, Germany\\
$^b$ Max-Planck Institut f\"ur
 Kernphysik, Postfach
103980,  69029 Heidelberg, Germany\\
$^c$ Joint Institute
 for Nuclear Research,
Dubna, 141980
 Moscow Region, Russia\\
\vspace*{1cm}
{{
\doublespace
\begin{quotation}
\noindent
{\bf Abstract:}
Interference between vector mesons 
electroproduced at different
 longitudinal coordinates
leads within Glauber approximation 
to a $Q^2$-dependence
of nuclear effects similar to what is 
expected to be an onset of color 
transparency (CT). We suggest such a 
mapping of the photon energies and virtualities, 
which eliminates the undesirable $Q^2$-variation
and allows to measure a net CT effect.

We develop a multichannel evolution equation,
which for the first time incorporates CT and 
the effects of the coherence length in the 
exclusive electroproduction of vector mesons.

\end{quotation}
\newpage

{\bf 1. Vector Meson production. Coherence
 length} 

 Vector mesons electroproduced at different points
 separated by longitudinal
distance $\Delta z$ have
 a relative phase shift $q_c\Delta z$, where 
$q_c = (Q^2 + m_V^2)/2\nu$ is the
longitudinal momentum transfer in 
$\gamma^*N \to VN$, 
$Q^2$ and $\nu$ are 
the virtuality and energy
 of the photon, respectively.
Taking this into account one arrives at the well known 
formula for nuclear transparency,
$Tr=\sigma_A/A\sigma_N$, for the coherent
 production of
vector mesons off nuclei 
 \cite{bauer},
 
 \beq
Tr_{coh} = \frac{(\stot)^2}{4A\sel}
 \int d^2b\left |\inf dz\
 \rho(b,z)\
e^{iq_cz}
 \exp\left[-{1\over 2}\stot\int_z^{\infty}
 dz'\rho(b,z')\right
]\right |^2
\label{1}
\eeq
 
 In the low-energy limit $q_c \gg 1/R_A$ and
 destructive
interference eliminates the coherent
 production. However,  nuclear transparency (\ref{1}) 
monotonically grows with
energy and reaches value $Tr_{coh} =
 \sigma_{el}^{VA}/A
\sigma_{el}^{VN}$. 
  Numerical examples for nuclear transparency in 
coherent production can be found in
 \cite{hkn1,kn1,hkn2}.
 
 Formula for
nuclear transparency for incoherent
 quasielastic photoproduction of vector
mesons in Glauber approximation  derived only
 recently \cite{hkn2}, reads
 
 \beqn
 & &Tr_{inc} =
\frac{\stot}{2A\sel}(\sin-\sel)
 \int d^2b\ \inf dz_2\ \rho(b,z_2)
\int_{-\infty}^{z_2} dz_1\ \rho(b,z_1)\
 \nonumber\\
 &\times &
e^{iq_L(z_2-z_1)}\
 \exp\left[ -{1\over 2}\stot\ \int_{z_1}^{z_2}dz
\rho(b,z)\right]
 \exp\left[ -\sin\ \int_{z_2}^{\infty} dz\
\rho(b,z)\right]\
 \nonumber\\
 &+&
 \frac{1}{A\sin}
 \int d^2b \left
[1-e^{-\sin T(b)}\right ]
 -Tr_{coh}\ ,
\label{2}
\eeqn
 
In contrast to coherent production nuclear transparency (\ref{2})
decreases with energy from $Tr_{inc}=
 \sigma_{in}^{VA}/A\sigma_{in}^{VN}$ at low energy 
($q_c \gg 1/R_A$) down to
value $Tr_{inc}=\sigma_{qel}^{VA}/A\sigma_{el}^{VN}$
at high energies ($q_c \ll 1/R_A$), where
$\sigma_{qel}^{VA}$ is the cross section of quasielastic 
$VA$ scattering.
Numerical examples are presented in \cite{hkn1,kn1,hkn2}.

Such an energy dependence is easily interpreted in terms 
of lifetime of the photon fluctuations, which propagate
over the coherence 
length $l_c \sim 1/q_c$: at low energy
($l_c \ll R_A$) the vector meson appears deep inside 
the nucleus and covers then about a half of the 
nuclear thickness. At high energy the vector meson preexists
as a photon fluctuation long time and propagates
through the whole nuclear thickness, resulting in a
more of absorption.

Variation of $l_c$ 
 may be caused by either its $\nu$- 
or $Q^2$-dependence. In the latter case $l_c$ decreases
with $Q^2$ and the nuclear transparency 
grows, what is expected to be a signature of color transparency (CT)
\cite{knnz}, a QCD phenomenon related to suppressed initial/final state
interaction in a nucleus for a 
small-size colorless wave packet \cite{zkl}. 
Examples for incoherent
electroproduction of $\rho$-meson on xenon are 
shown in Fig.~1 (more examples are 
in \cite{hkn1,kn1,hkn2}). The predicted  $Q^2$-dependence
is so steep that makes it quite problematic to observe an additional
$Q^2$-dependence generated by the color transparency effects.

Note that the cross section of photoproduction
of the vector mesons on nuclei is energy-dependent at low energy
due to quark (Reggeon) exchanges. 
This is very easy to include 
in our calculations, but we 
neglect the energy-dependence for the sake of simplicity.
\medskip

\noi
{\bf 2. Beyond Glauber approximation. Formation
length.} 
\medskip

According to the uncertainty principle one needs (formation) 
time to resolve
different levels, $V,\ V'...$, in the final state. 
Corresponding formation length 
$l_f \approx 2\nu/(m^2_{V'}
- m_V^2)$ is longer than the coherence length.
It has a close relation to the onset of CT \cite{kz,knnz}, 
which is possible only
if $l_f \geq R_A$. Note that a full CT effect at
$Q^2 \gg m_V^2$ imposes a stronger condition $l_fm_V/\sqrt{Q^2} 
\gg R_A$.  This is because
a small size, $\sim 1/Q^2$, wave packet, decomposed over different 
$V$-meson states, includes all states up to $M^2 \sim Q^2$.
Such a decomposition must be frozen by Lorentz time dilation
up to the heaviest states, what leads to above condition within
oscillatory model.

At medium energies and $Q^2$ one can expect only an onset of CT,
i.e. a monotonic growth of $Tr(Q^2)$ with $Q^2$, which
results mostly from interference of the two lowest states,
$V$ and $V'$.
Inclusion of the second channel in the case of coherent 
production is quite simple and the results are presented in \cite{kn1}.
We concentrate on the incoherent electroproduction here.
For the first time we present the evolution equation 
and its solution incorporating both the coherence and formation
length effects, what is of a special importance in the
intermediate energy range. 

The solution of the general multichannel equation
will be presented elsewhere. Here we consider 
the two-coupled-channel case. Since we have to sum over
all final states of the nucleus, the
propagation of the charmonium wave packet through
 the nucleus is to be
described by density matrix $P_{ij} = 
\sum |\psi_i\ra|\psi_j\ra^+$.
The wave function $|\psi_i\ra$ has three components,
$\gamma^*$, $V$ and $V'$.
The evolution equation reads,

 \beq
i\frac{d}{dz}\widehat P =
 \widehat Q \widehat P -
 \widehat P \widehat Q^+ -
 {i\over 2}\sigma^{V N}_{tot}
 \left(\widehat T \widehat P +
 \widehat
P \widehat T^+\right) +
 i\sigma^{VN}_{el}
 \widehat T \widehat P
\widehat T^+.\ 
\label{11}
\eeq
\noi
Here
 \beq
 \widehat Q =
 \left(\begin{array}
{ccc}0&0&0\\0&q&0\\0&0&q'
\end{array}\right)\ ;\ \ \ \ 
 \widehat T =
 \left(\begin{array}
 {ccc}0&0&0\\\lambda&1&\epsilon
 \\\lambda
R&\epsilon&r
\end{array}\right)
\label{8}
\eeq
\noi
are the longitudinal momentum transfer 
and scattering amplitude
operators, respectively,
where $q(q') = (m^2_{V(V')} + Q^2)/2\nu$. 
For other parameters we use notations from
\cite{hk}, $r=\sigma_{tot}^{V'N}/\sigma_{tot}^{VN}$,
$\epsilon = f(VN \to V'N)/f(VN \to VN)$ and
$R= f(\gamma N \to V'N)/f(\gamma N \to VN)$.
The value of parameter $\lambda = 
f(\gamma N \to V'N)/f(VN \to VN)$
is inessential, since it cancels in nuclear transparency.
The boundary condition for the density matrix is
$P_{ij}(z\to -\infty) = \delta_{i0}\delta_{j0}$.
Note, the same eq.~(\ref{11}) reproduces Glauber approximation
(\ref{2}) when $\epsilon=0$. It also calculates coherent production 
when one fixes $\sigma_{el}^{VN} = 0$.

Combination of the two effects of coherence and formation (onset of CT)
lengths may provide an unusual energy behaviour of nuclear transparency.
An example is shown in Fig.~2 for  real photoproduction of
$\rho$ and $\rho'$ on lead predicted by eq.~(\ref{11}).
We try two models for the $\rho$ and $\rho'$ wave functions,
the nonrelativistic oscillator and a relativized version:
i) $r=1.5,\ \epsilon=-0.5,\ R^2=0.074$ ($Q^2=0$) for the nonrelativistic
oscillator; ii) $r=1.25,\ \epsilon=-0.14,\ R^2=0.22$ for the relativized 
wave function \cite{kn1}. We calculate the $Q^2$-dependence of
$R^2$ in accordance with \cite{kn1}.
Our predictions for energy dependence of nuclear 
transparency for incoherent photoproduction of $\rho$ 
and $\rho'$ are shown in Fig.~2 for the two sets of
parameters. While
both variants give similar results for the $\rho$, we expect
a nontrivial energy dependence for the $\rho'$, which results 
from interplay of the effects of coherence and formation lengths.
It turns out to be 
extremely sensitive to the form of the wave function.
Thus, such kind of measurement may bring unique information about
the structure of vector mesons.

}}
\begin{figure}[tbh]
 \includegraphics{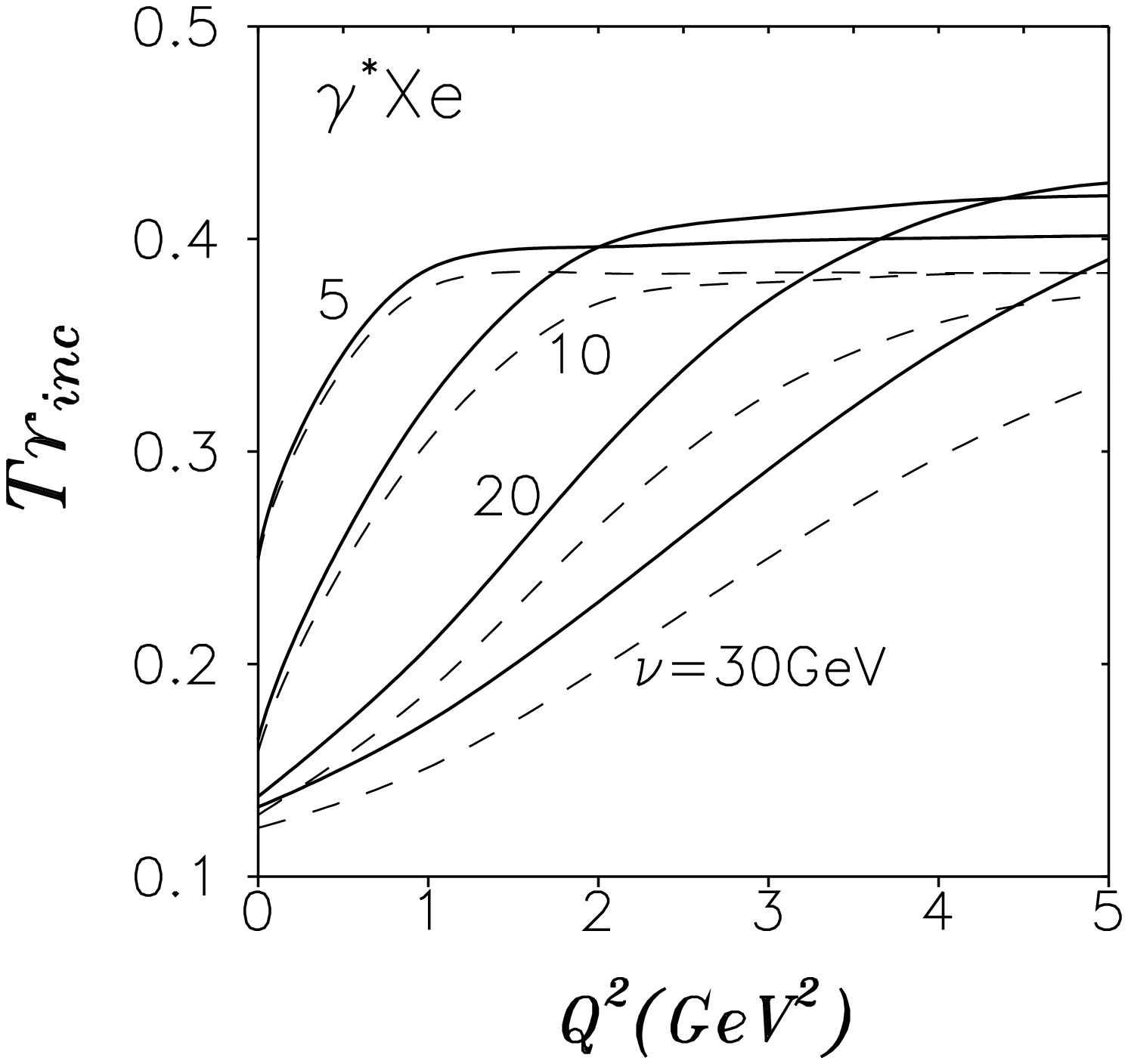}
\includegraphics{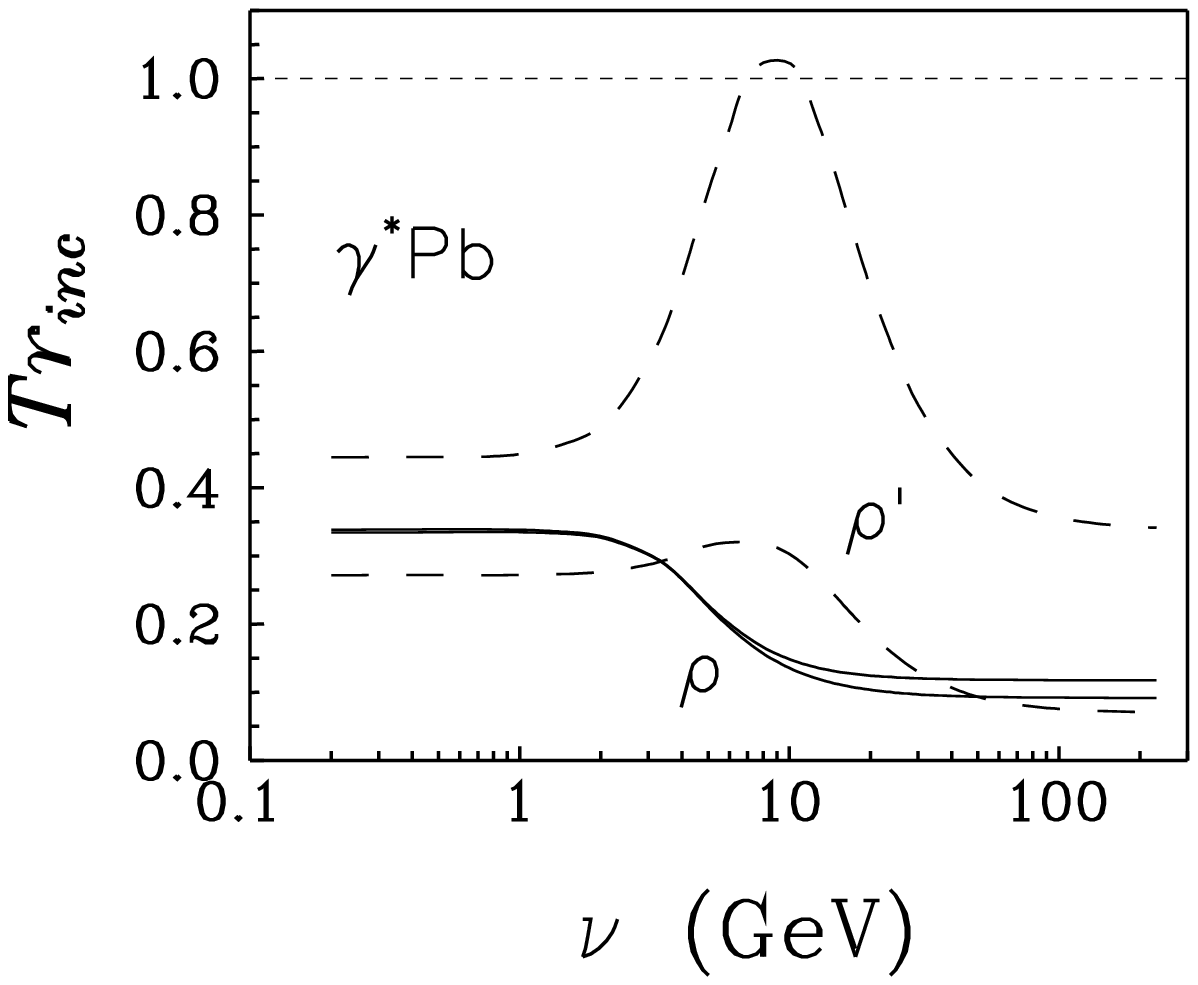}
\begin{center}
\vspace{7cm}
\parbox{16cm}
 {\caption[Delta]
 {\it $Q^2$-dependence of nuclear transparency for
$\rho$-meson electroproduction on xenon at
photon energies $\nu = 5,\ 10, 20$ and $30\ GeV$.
Dashed curves correspond to Glauber approximation,
solid curves are calculated with the evolution 
equation eq.~(\ref{11}).}
\label{fig1}}
\parbox{16cm}
{\caption[Delta]
 {\it Nuclear transparency versus $\nu$ for real
photoproduction of $\rho$ (solid curves) and $\rho'$ 
(dashed curves) on lead. Upper curves correspond to 
the nonrelativistic oscillatory wave functions, while
the bottom curves are calculated with relativized variant
\cite{nnz}}
\label{fig2}}
\end{center}
\end{figure}

\begin{figure}[tbh]
 \includegraphics{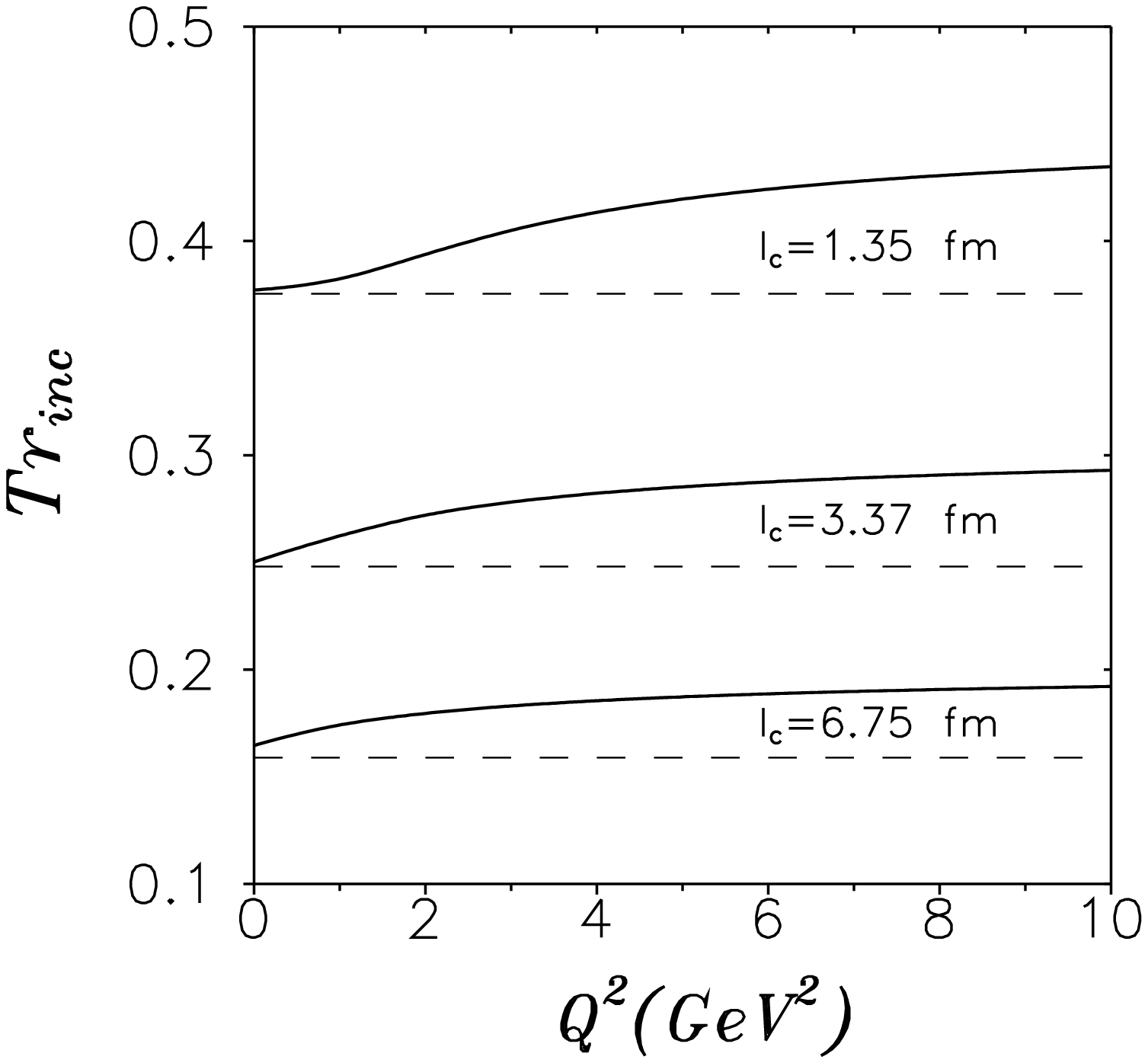}
\includegraphics{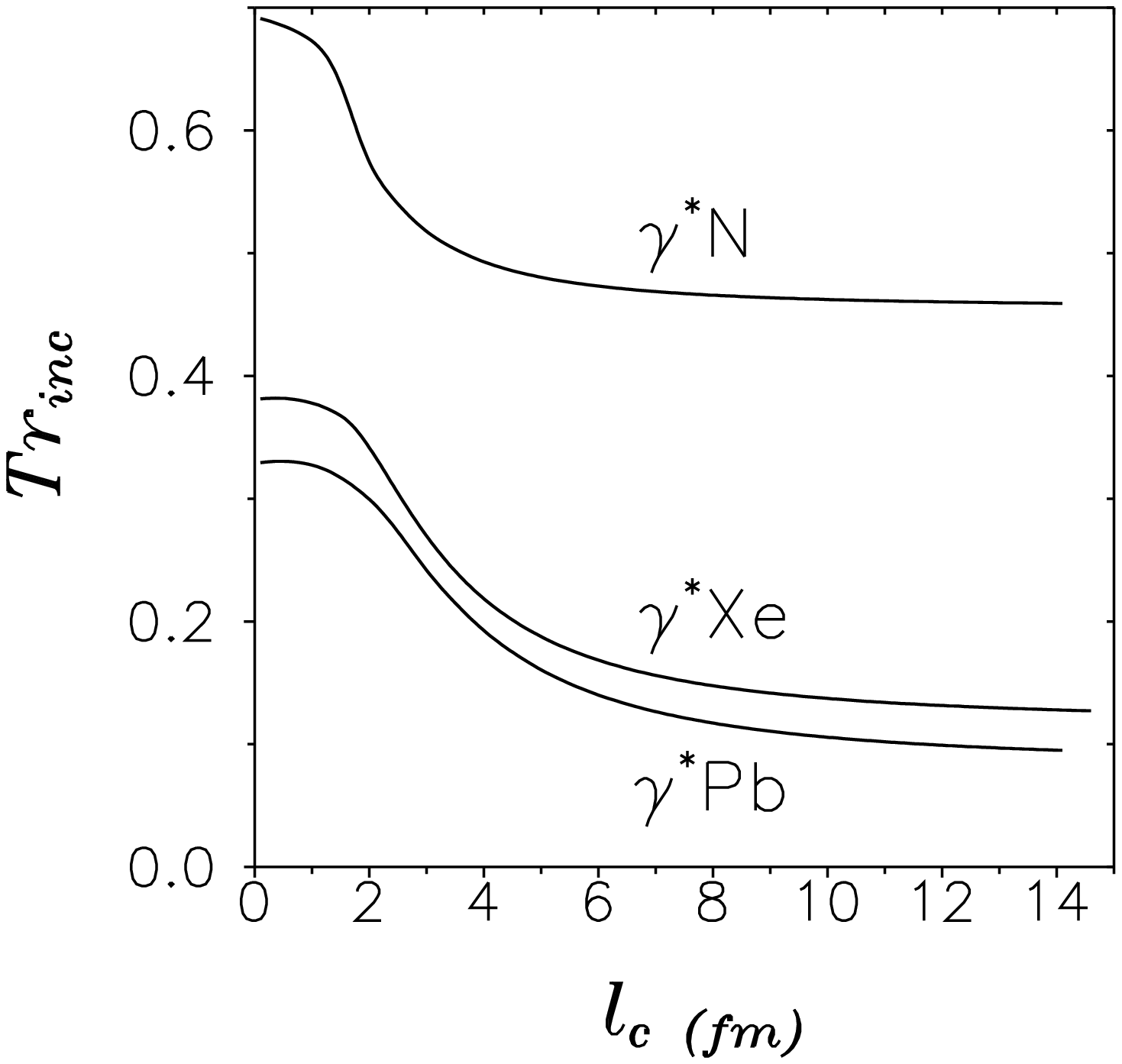}
\begin{center}
\vspace{6.5cm}
\parbox{16cm}
 {\caption[Delta]
 {\it $Q^2$-dependence of nuclear transparency
at fixed $l_c=const$ for xenon. 
Glauber approximation
predictions are shown by dashed curves.
Solid curves demonstrate the effect of CT in 
the two coupled channel approximation}
\label{fig3}}
\parbox{16cm}
{\caption[Delta]
 {\it Nuclear transparency for nitrogen, xenon and lead 
versus the coherence length in
Glauber approximation.}
\label{fig4}}
\end{center}
\end{figure}

 {{
\doublespace

Note that at low energy, where the eikonal approximation
is exact, the nuclear transparency for the $\rho'$ is about the same
as for the $\rho$ despite the bigger radius and stronger 
absorption of the $\rho'$. This is because the transitions
$\rho \rightleftharpoons \rho'$ are possible even without 
interference, but their probability is $\propto 
\sigma_{el}^{VN}$ rather than $\sigma_{tot}^{VN}$. For this
reason it is very much suppressed for $J/\Psi$, but not for 
$\rho$.

We have arrived at the main point: how to search for a signal
of CT? The two coupled channels are supposed to reproduce well
the onset of CT, which is expected to manifest itself 
as an additional growth of
nuclear transparency with $Q^2$. The results obtained with
eq.~{11} and relativized wave functions are shown in
Fig.~1 by solid curves. We see that variation of the transparency
with $Q^2$ 
due to the coherence length effect predicted
by Glauber model is rather steep and makes it very difficult
to detect a CT signal even with high-statistics measurements.

In order to single out the effect of CT one should remove
the $l_c$-dependence of nuclear transparency. This can be
done mapping values of $Q^2$ and
$\nu$ in a way, which keeps $l_c=2\nu/(m_V^2+Q^2)$ constant.
Our predictions for $Q^2$-dependence of nuclear transparency
at different values of $l_c$ are plotted in Fig.~3 both
for single- (Glauber) and double-channel approaches.
The former, as we expected, provides a constant nuclear
transparency, while the latter results in a rising $Tr(Q^2)$. 
Observation of
such a dependence on $Q^2$ should be grated as an onset of CT.
We show $l_c$-dependence of transparency calculated in 
Glauber approximation in Fig.~4.
\medskip

\noi
{\bf 3. Onset of CT in $(e,e'p)$}
\medskip

The principal difference of $(e,e'p)$ reactions from 
the above discussed photoproduction of vector mesons is
a strong correlation between the values of $Q^2$ and
the energy of the recoil proton, $E_p \approx m_p +
Q^2/2m_p$. Therefore, while one wants just to increase the proton
energy in order to freeze its small size, one unavoidable
has to increase $Q^2$, i.e. to suppress the cross section.
This is why no CT effect is still detected in this reaction.
As was demonstrated above, in order to freeze size
$\sim 1/Q^2$ of the ejectile during propagation through 
the nucleus, its energy must be
$\nu > R_A\sqrt{Q^2}\omega$, where $\omega \approx 0.3\ GeV$ is
the oscillatory parameter. This demands unreachable values of
$Q^2$ of a few hundred $GeV^2$. The same problem concerns 
$(p,2p)$ quasielastic proton scattering at large angles.

On the other hand, a growth of $Q^2$ causes
an increase of the ejectile energy. It is known, however,
since \cite{gribov} that nuclear matter is more transparent
at higher energies due to inelastic shadowing corrections.
This effect is general and independent of color dynamics or 
whether the CT effect
exists or not. The growth of nuclear transparency in $(e,e'p)$
with energy, and consequently with $Q^2$, was estimated in 
\cite{kn2}, assuming a simplest ''anti-CT'' scenario: a regular
proton, rather than a small-size fluctuation, 
is knocked out in $(e,e'p)$ reaction. The predicted $Q^2$
dependence turns out to be similar to what one expects to be
an onset of CT up to about $Q^2 \approx 20\ GeV^2$.
Therefore, one should be very cautious interpreting the growth
of $Tr(Q^2)$ as a signal of CT in this region. It can be safely 
done only at quite high, still unreachable $Q^2$.
\medskip

\noi
{\bf 4. Conclusion} 
\medskip

We have developed for the first time a multichannel
approach to incoherent exclusive electroproduction of
vector mesons off nuclei, which incorporated effect of
the coherent and formation lengths. It is based on a multichannel evolution
equation for the density matrix of the produced wave packet.
Variation of the coherence length with the photon energy and $Q^2$ 
causes substantial changes of the nuclear transparency even in
Glauber approximation. This fact makes it
very difficult to observe an onset of CT. We suggest such a mapping 
of $\nu$ and $Q^2$ values, which keeps the coherence length constant.
in  This case
the nuclear transparency cannot change within Glauber approximation.
We provide estimates of the onset of CT within two coupled channel
model.

One faces even more problems searching for a signal of CT in
$(e,e'p)$ quasielastic scattering, due to its specific kinematics.
It is very difficult to interpret the growth of nuclear transparency with $Q^2$ 
as an onset of CT at $Q^2 < 20\ GeV^2$.

}}
\setlength{\baselineskip} {5pt}


\begin{thebibliography}{MMM}
\bibitem{bauer} T.H.~Bauer et al., Rev. Mod. Phys. {\bf 50} (1978) 261
\bibitem{hkn1} J.~H\"ufner, B.Z.~Kopeliovich and J.~Nemchik,
in the Proceedings of ELFE Workshop, July 1995, ed. S.D.~Bass,
hep-ph/9511215
\bibitem{kn1} B.Z.~Kopeliovich and J.~Nemchik, MPIH-V41-1995, 
nucl-th/9511018
\bibitem{hkn2} J.~H\"ufner, B.Z.~Kopeliovich and J.~Nemchik,
DOE/ER/40561-260-INT96-19-13, nucl-th/9605007, to appear in Phys. Lett. B
\bibitem{knnz} B.Z.~Kopeliovich, J.~Nemchik, N.N.~Nikolaev and B.G.~Zakharov,
Phys. Lett {\bf B324} (1994) 469; Phys. Lett. {\bf B309} (1993) 179
\bibitem{zkl} Al.B.~Zamolodchikov, B.Z.~Kopeliovich and L.I.~Lapidus,
 Sov. Phys. JETP Lett. {\bf 33}, (1981) 612
\bibitem{kz} B.Z.~Kopeliovich and B.G.~Zakharov, Phys. Rev. {\bf
D44} (1991) 3466
\bibitem{hk} J.~H\"ufner and B.Z.~Kopeliovich,
Phys. Rev. Lett. {\bf 76} (1996) 192
\bibitem{nnz} J.~Nemchik, N.N.~Nikolaev, E.~Predazzi and B.G.~Zakharov,
DFTT 71/95, hep-ph/9605231, to appear in Z. Phys.C
\bibitem{gribov} V.N.~Gribov, Sov. Phys. JETP, {\bf 29} (1969) 483
\bibitem{kn2} B.Z.~Kopeliovich and J.~Nemchik, Phys. Lett. {\bf 
B 368} (1996) 187

\end{thebibliography}
\end{document}